\documentclass[aps,prd,twocolumn]{revtex4}

\usepackage{latexsym}
\usepackage{amssymb}
\usepackage{psfrag}
\usepackage{graphicx}

\newcommand \g{\gamma}
\newcommand \p{\partial}

\newcommand \D{D}

\begin{document}

\title{Constraint damping in the Z4 formulation and harmonic gauge}

\author{Carsten Gundlach}
\author{Gioel Calabrese}
\author{Ian Hinder}
\affiliation{School of Mathematics, University of Southampton,
         Southampton SO17 1BJ, UK}

\author{Jos\'e M. Mart\'\i n-Garc\'\i a}
\affiliation{Instituto de Estructura de la Materia, Centro de F\'\i sica
Miguel A. Catal\'an, C.S.I.C., Serrano 123, 28006 Madrid, Spain}

%%%%%%%%%%%%%%%%%%%%%%%%%%%%%%%%%%%%%%%%%%%%%%%%%%%%%%%%%%%%%%%%%%%%%%%%

\begin{abstract}

We show that by adding suitable lower-order terms to the Z4
formulation of the Einstein equations, all constraint violations
except constant modes are damped. This makes the Z4 formulation a
particularly simple example of a $\lambda$-system as suggested by
Brodbeck et al. We also show that the Einstein equations in harmonic
coordinates can be obtained from the Z4 formulation by a change of
variables that leaves the implied constraint evolution system
unchanged. Therefore the same method can be used to damp all
constraints in the Einstein equations in harmonic gauge. 

\end{abstract}

%%%%%%%%%%%%%%%%%%%%%%%%%%%%%%%%%%%%%%%%%%%%%%%%%%%%%%%%%%%%%%%%%%%%%%%%

\maketitle

%%%%%%%%%%%%%%%%%%%%%%%%%%%%%%%%%%%%%%%%%%%%%%%%%%%%%%%%%%%%%%%%%%%%%%%%

\section{Introduction}

Formulations of the Einstein equations as an initial-boundary value
problem are required for the numerical simulation of astrophysical
events, such as the inspiral and merger of a binary system of black
holes. Simulations are still crucially limited by instabilities. Most
of these instabilities arise already in the continuum system of
partial differential equations, rather that at the stage of finite
differencing.

There is now broad agreement that in order to avoid some of these
instabilities, the formulation of the Einstein equations that one uses
should give rise to well-posed problems. For the Cauchy problem
(no boundaries in space) to be well-posed, strong hyperbolicity is a
necessary and sufficient criterion. Symmetric hyperbolicity implies
strong hyperbolicity and allows one to formulate a well-posed
initial-boundary value problem. Symmetric hyperbolicity of the
subsidiary constraint evolution system is likely to be a crucial
ingredient in making the boundary conditions consistent with the
constraints. 

In a hyperbolic formulation, the error associated with constraint
violation grows at a bounded rate, but this can be very fast in
practice. It would be preferable if one could find a formulation of
the Einstein equations in which the submanifold of solutions that also
obey the constraints is an attractor. Clearly this requires a
mechanism for breaking the time-reversal symmetry of general
relativity away from the constraint surface. Mechanisms that have been
suggested include dynamically adjusting the free parameters of the
constraint addition \cite{Tiglio}, or adding derivatives of the
constraints so that the system becomes mixed parabolic and hyperbolic
\cite{Calabrese}.

Brodbeck et al \cite{Brodbecketal} have suggested a general approach
called the $\lambda$-system to solving a system of evolution equations
and constraints such that the constraint surface is an attractor. This
consists in adding one variable $\lambda$ for each constraint, such
that the time derivatives of the $\lambda$ are the constraints, and
the extended system is (or remains) symmetric hyperbolic. One then
adds damping terms $\p_t \lambda=\dots - \kappa \lambda$ to the
evolution equations for the variables $\lambda$. As they are lower
order, they do not affect the hyperbolicity of the system. These terms
should damp the $\lambda$, and therefore the constraints. 

In \cite{Brodbecketal} the Frittelli-Reula symmetric hyperbolic
formulation of the Einstein equations was extended in this way, and it
was shown analytically that when the system is linearised around
Minkowski spacetime, the constraint surface is an attractor. In
\cite{SiebelHuebner} the conformal field equations, reduced by two
Killing vectors, were extended in the same way, and investigated
numerically. This worked well for linear gravity, but not in strong
field tests, where the constraint violations were reduced, but not to
zero, and the error actually increased.

%%%%%%%%%%%%%%%%%%%%%%%%%%%%%%%%%%%%%%%%%%%%%%%%%%%%%%%%%%%%%%%%%%%%%%%%

\section{Z4 as a simple $\lambda$-system}

Our starting point is the observation that another symmetric
hyperbolic formulation of the Einstein equations, the Z4 formulation
\cite{Z4I,Z4II,Z4III} is already a $\lambda$ system, without the need
to add extra variables. (This was already noted in \cite{Z3} for the
Z3 system, which is closely related to the Z4 system.) For simplicity
we restrict our presentation to the vacuum case. Including matter
would be straightforward.

The Z4 system is obtained in its 4-dimensional covariant form by
replacing the vacuum Einstein equations $R_{ab}=0$ by
\begin{equation}
\label{Z4}
R_{ab}+\nabla_a Z_b +\nabla_b Z_a=0, 
\end{equation}
where $R_{ab}$ is the Ricci tensor of the 4-dimensional spacetime
metric $g_{ab}$ and $Z_a$ is an additional vector field. The main
effect of this extension is to turn the 4 Einstein constraints into
first-order evolution equations for the 4-vector $Z_a$.  A solution of
the Z4 equations is a solution of the Einstein equations if and
only if $Z_a=0$. (There is one exception: if the spacetime admits
a Killing vector, a solution of the Z4 equations with $Z_a$ equal to
the Killing vector is also a solution of the Einstein equations. In
the following, we assume for simplicity that the spacetime is generic
and does not admit a Killing vector.)

We shall see that the variables $Z_a$ are already variables of the
$\lambda$ type. All we need to do is to add the damping term. We do
this in covariant notation by replacing the Einstein equations by
\begin{equation}
\label{Ricci}
R_{ab}+\nabla_a Z_b +\nabla_b Z_a 
-\kappa\left[t_a Z_b+t_b Z_a-(1+\rho)g_{ab}t^c Z_c\right]=0, 
\end{equation}
or, with the trace reversed,
\begin{equation}
\label{Einstein}
G_{ab}+\nabla_a Z_b +\nabla_b Z_a 
- g_{ab}\nabla^c Z_c 
-\kappa(t_a Z_b+t_b Z_a+\rho g_{ab}t^cZ_c)=0 , 
\end{equation}
where $t^a$ is a non-vanishing timelike vector field and $\kappa\ge
0$ and $\rho$ are real constants. (We shall later restrict to
$\rho=0$.) 
It is $t^a$ that explicitly breaks time reversal
invariance. A simple geometrical choice is $t^a=n^a$, the future
pointing unit normal vector on the time slicing. 

We carry out the usual 3+1 split of the metric as
\begin{equation}
ds^2=-\alpha^2\, dt^2+\gamma_{ij}(dx^i+\beta^i\,dt)(dx^j+\beta^j\,dt),
\end{equation}
and split $Z_a$ as $Z^a=X^a+n^a\theta$ where 
$\theta\equiv -n_a Z^a$. In adapted coordinates this gives
$\theta=\alpha Z^0$ and $X_i=Z_i$. (Note that $n^a$ as defined in
\cite{Z4I} is past-pointing, while ours is future-pointing, so that
the two definitions of $\theta$ are the same.) In the following we use
the established, but slightly ambiguous, notation $Z_i$ to denote $X_i$.  We
denote the Ricci tensor of $\gamma_{ij}$ by $R_{ij}$ and its covariant
derivative by $D_i$. Spatial tensor indices $i,j$ are moved with
$\gamma_{ij}$. We also define the derivative operator $\p_0\equiv
\alpha^{-1}(\partial_t -{\cal L}_\beta)$. In this notation, the 3+1
split of the Z4 equations is
\begin{eqnarray}
\label{d0gij}
\p_0 \g_{ij}&=&-2K_{ij}, \\
\label{d0Kij}
\p_0 K_{ij}&=& -\alpha^{-1}D_iD_j\alpha+R_{ij}-2K_{ik}{K^k}_j+K K_{ij}
\nonumber \\
&&+\D_i Z_j + \D_j Z_i - 2\theta K_{ij} 
-\kappa(1+\rho)\g_{ij}\theta, \\
\label{d0theta}
\p_0 \theta &=& {1\over 2} H - \theta K + \D_k Z^k - \D_k(\ln
\alpha) Z^k \nonumber \\
&&-(2+\rho)\kappa\theta, \\
\label{d0Xi}
\p_0 Z_i &=& M_i + \D_i \theta - \D_i(\ln\alpha) \theta -
2{K_i}^k Z_k
-\kappa Z_i.
\end{eqnarray}
In the last two equations $H$ and $M_i$ are shorthand for the Einstein
constraints
\begin{eqnarray}
H&\equiv &R-K_{ij}K^{ij}+K^2, \\ 
M_i&\equiv&D^j K_{ji}-D_i K.
\end{eqnarray}

What happens when the initial data for $\g_{ij}$ and $K_{ij}$ do
not obey the Einstein constraints? By substituting the definitions
of $H$ and $M_i$ into the evolution equations
(\ref{d0gij})-(\ref{d0Kij}), we find their formal time derivatives
\begin{eqnarray}
\label{d0H}
\partial_0 H &=& 
- 4 M^i D_i(\ln\alpha) 
+ 2 K H 
- 2 D^i M_i 
-4\kappa (1+\rho)K\theta\nonumber\\
&& +K_{ij} (\g^{ij}\g^{kl}-\g^{ik}\g^{jl})
[4D_lZ_k-4\theta K_{kl}]  , \\
\label{d0Mi}
\partial_0 M_i &=& 
-\frac{1}{2}D_i H
+(K-2\theta) M_i 
- D_i(\ln\alpha) H 
+ 2 R_{ij} Z^j 
\nonumber \\ &&
+ D^j(D_j Z_i-D_i Z_j)
+\alpha^{-1}D^j\alpha (D_iZ_j+D_jZ_i)
\nonumber \\ && 
- 2D_i(\ln\alpha) D^j Z_j 
-2\alpha^{-1}K_{ij}D^j(\alpha\theta)
\nonumber \\ &&
+ 2\left[K +\kappa(1+\rho)\right]
\alpha^{-1} D_i(\alpha\theta) .
\end{eqnarray}
In order for the constraint evolution system to close (in the usual
sense, see below) it must be considered to consist of
(\ref{d0H},\ref{d0Mi},\ref{d0theta},\ref{d0Xi}) for the constraint
variables $H$, $M_i$, $\theta$ and $Z_i$. In particular, a
solution of the evolution equations obeys $H=M_i=\theta=Z_i=0$ at all
times if and only if they all vanish at $t=0$. The constraint system
associated with the Z4 system is unusual in that $\theta$ and $Z_i$
are genuine dynamic variables while, as in other formulations, $H$ and
$M_i$ are only shorthands for combinations of the dynamic variables
$\g_{ij}$ and $K_{ij}$ and their derivatives.

One can replace the 8 first-order evolution equations of the
constraint system by a second-order wave equation for the
4-vector $Z_a$ by taking the divergence of (\ref{Einstein}), and using
the contracted Bianchi identity $\nabla^a G_{ab}=0$.  The result is
\begin{equation}
\label{Zwave}
\Box Z_b+R_{ab}Z^a -\kappa
\nabla^a \left(t_a Z_b+t_b Z_a+\rho g_{ab} t^c Z_c\right)=0,
\end{equation}
where $\Box$ is the covariant wave operator $\nabla_a\nabla^a$. Given
$Z_a$, the Einstein constraints $G^{0\mu}$ can be read off from
(\ref{Einstein}), or in 3+1 form $H$ and $M_i$ can be read off from
(\ref{d0theta})-(\ref{d0Xi}). Note that for $Z_i=\theta=0$ at some
instant $t=0$, the condition $\dot{Z_i}=\dot{\theta}=0$ is equivalent to
$H=M_i=0$ at that instant.  That means that all four constraints vanish
at all times if and only if $Z_i=\theta=\dot{Z_i}=\dot{\theta}=0$ at
$t=0$.

In either its first-order or second-order form, the constraint
evolution system closes only in the sense that the right-hand side is
proportional to the constraints, but not in the sense that it is
autonomous: one cannot consistently evolve either (\ref{Zwave}) or
(\ref{d0H},\ref{d0Mi},\ref{d0theta},\ref{d0Xi}) while considering the
variables $\g_{ij}$ and $K_{ij}$ (or equivalently the spacetime metric
$g_{ab}$) as fixed. Instead one should focus on the following
question:

When one evolves initial data with a small constraint violation (set for
example by finite differencing error) does the constraint violation
grow or decay as the initial data are evolved? To address this, we perturb
around a background solution $g_{ab}^{(0)}$ that obeys all 10 Einstein
equations, and write
\begin{equation}
g_{ab}={}g_{ab}^{(0)}+\epsilon\,g_{ab}^{(1)},
\end{equation}
where $R_{ab}^{(0)}= 0$ and
$Z_a^{(0)}=0$. To first order in $\epsilon$ the constraint violation
then obeys a linear evolution equation with coefficients taken from
the background Einstein spacetime, and admitting arbitrary data. In
the Z4 formulation this is just a vector wave equation on the
background spacetime, namely
\begin{equation}
\label{Zwave1}
\Box^{(0)} Z_b^{(1)}-\kappa
\nabla^{a(0)}\left(t_a^{(0)}  Z_b^{(1)}
+t_b^{(0)}  Z_a^{(1)}+\rho g_{ab}^{(0)}t^{c(0)}Z_c^{(1)}\right)=0.
\end{equation}
The growth of sufficiently small constraint violations is controlled
by the damping; for larger constraint violations nonlinear lower-order
terms also become important. We expect that with sufficiently large
$\kappa$, and starting from sufficiently small initial constraint
violations, the nonlinear terms never become important in practice.

[Note added in revision: Friedrich \cite{Friedrich} has has
independently carried out a partial non-linear analysis of
(\ref{Zwave}) in the context of generalised harmonic coordinates and
without damping. In (\ref{Zwave}) he considers the spacetime in the
wave operator $\Box$ as given, but expresses $R_{ab}$ in terms of
$Z_a$ using (\ref{Ricci}). The result is wave equation for $Z_a$ with
a quadratic lower-order term, but on a fixed spacetime. He shows that
generic solutions blow up in finite time. This analysis accounts for
some of the hidden nonlinearity of (\ref{Zwave}), but does not include
all terms of $O(\epsilon^2)$. We have restricted ourselves to a
consistent linear analysis.]

%%%%%%%%%%%%%%%%%%%%%%%%%%%%%%%%%%%%%%%%%%%%%%%%%%%%%%%%%%%%%%%%%%%%%%%%

\section{Mode analysis}

The constraint evolution system of the Z4 system is simpler than in
other formulations of the Einstein equations in that it has the form
of a covariant wave equation. We can use this to carry out a mode
analysis for the linearised constraint system (\ref{Zwave1}) on an
arbitrary Einstein background in the high-frequency, frozen
coefficient limit in which the wavelength of $Z_a^{(1)}$ is much
smaller than the curvature scale of the Einstein background. We can
then locally approximate the background $g_{ab}^{(0)}$ as Minkowski
space, and work in standard Minkowski coordinates. Assuming
without loss of generality that $t_at^a=-1$, we go to the frame in
which $t^\mu=(1,0,0,0)$ and find (now dropping the expansion indices)
\begin{eqnarray}
\Box Z_0-\kappa\left[(2+\rho) \p_t Z_0-\p^i Z_i\right]&=&0, \\
\Box Z_i-\kappa\left(\p_t Z_i+\rho \p_i Z_0\right)&=&0,
\end{eqnarray}
where $\Box$ is now the Minkowski wave operator
$-\partial_t^2+\partial_i\partial^i$.
We make a plane-wave ansatz
\begin{equation}
Z_\mu(t,x^i)=e^{st+i\omega_i x^i} \hat Z_\mu, 
\end{equation}
with complex $s$ and real $\omega_i$. This gives rise to the
eigenvalue problem
\begin{widetext}
\begin{equation}
\left(\begin{array}{ccc}
-s^2-\omega^2-\kappa(2+\rho)s & \kappa i\omega & 0 \\
-\kappa\rho i\omega & -s^2-\omega^2-\kappa s & 0 \\
0 & 0 & -s^2-\omega^2 -\kappa s 
\end{array}\right)
\left(\begin{array}{c}
\hat Z_0 \\ \hat Z_n \\ \hat Z_A
\end{array}\right)=0.
\end{equation}
\end{widetext}
where $Z_n$ is the component of $Z_i$ in the direction of $\omega_i$
and $Z_A$ stands for the projection of $Z_i$ normal to $\omega_i$. 
The 4 eigenvalues $s$ for $Z_A$ are
\begin{equation}
\label{ZA_s}
s=-{\kappa\over2}\pm\sqrt{\left({\kappa\over 2}\right)^2-\omega^2}
\end{equation}
(each of these occurring twice),
independently of $\rho$. The other 4 eigenvalues $s$ are in general
complicated. However, for $\rho=0$ they take the simple form
\begin{equation}
\label{Z0n_s}
s=-\kappa\pm\sqrt{\kappa^2-\omega^2}
\end{equation}
(each of these occurring twice).  Therefore with $\rho=0$ all modes are
damped for all $\omega_i\not=0$. At high wave numbers, $\omega\gg
\kappa$, the damping is by a constant factor, with
\begin{equation}
s\simeq -\kappa\pm i\omega, -{\kappa\over 2}\pm i\omega,
\end{equation}
but at low wave numbers, $\omega\ll\kappa$, it is similar to a heat
equation, with
\begin{equation}
s \simeq -\kappa,-{\omega^2\over \kappa},-2\kappa,-{\omega^2\over 2\kappa}.
\end{equation}
Half of the modes are damped less with decreasing wavenumber, and not
at all at zero wavenumber. The only other case in which the
eigenvalues $s$ are simple is $\rho=-1$. This also simplifies the
field equations, but in this case the other 4 eigenvalues are
$s=-\kappa\pm i\omega$ and $s=\pm i\omega$, and so there are undamped
modes for all $\omega_i$. It turns out that all but the constant modes
are damped for any $\rho>-1$, but $\rho=0$ is the most natural choice,
and we assume this value in the following.

We have shown that the constraint manifold is an attractor for
$\kappa>0$ when the equations are linearised around Minkowski space,
with the exception of constraint violations that are constant in
space. The same is true in the high-frequency limit when linearising
around any Einstein background. This analysis breaks down when the
constraint violations are large and/or when their wavelength is large
compared to the background curvature. In that case lower-order terms
can potentially make the constraints grow against the explicit damping
term. However, compared to the systems of
\cite{Brodbecketal,SiebelHuebner}, $Z4$ has only a fraction of the
number of variables and constraints, and so is less likely to be
affected by undesirable growth arising from lower order terms.

%%%%%%%%%%%%%%%%%%%%%%%%%%%%%%%%%%%%%%%%%%%%%%%%%%%%%%%%%%%%%%%%%%%%%%%%

\section{Z4 and harmonic gauge}

The hyperbolicity of any formulation of the Einstein equation depends
on the choice of gauge (assuming that the formulation does not already
fix the gauge). Formulations of the Einstein equations derived from
the ADM formulation are typically not hyperbolic when the lapse $\alpha$
and shift $\beta^i$ are given functions of the coordinates, but if
they can be made hyperbolic at all, this is typically true for fixed
shift and fixed densitised lapse $Q$, where $\alpha$ and $Q$ are
related by 
\begin{equation}
\label{densitisedlapse}
\alpha \equiv \gamma^{\sigma/2} Q,
\end{equation}
for some constant parameter $\sigma$. The Z4 formulation with fixed
densitised lapse and fixed shift is strongly hyperbolic for
$\sigma>0$. Surprisingly, it is not symmetric hyperbolic for any
$\sigma$. (The most general energy that is conserved in the
high-frequency, frozen coefficient approximation fails to be positive
definite. Details will be given elsewhere.)

An evolved version of the densitised lapse (usually called the
Bona-Mass\'o lapse) is 
\begin{equation}
\label{liveslicing}
\p_0 \ln \alpha = -\sigma(K-m\theta) + \p_0 \ln Q,
\end{equation}
where $m$ is another constant parameter. When $\theta=0$, the $\p_0$
derivative of (\ref{densitisedlapse}) is just (\ref{liveslicing}). As
pointed out in \cite{Z4I}, the Z4 formulation is symmetric hyperbolic
with the lapse (\ref{liveslicing}) and fixed shift for $\sigma=1$ and
$m=2$. For any other $\sigma>0$, it is strongly hyperbolic (with
arbitrary $m$), but not symmetric hyperbolic. (The most general energy
fails to be positive definite for $\sigma\ne 1$. Details will be given
elsewhere.)  $\sigma=1$ is equivalent to harmonic slicing. This
suggests that in some way Z4 is closely related to the Einstein
equations in harmonic coordinates.

In fact, if in (\ref{Z4}) we consider $Z_\mu$ as the shorthand
\begin{equation}
\label{H4}
Z_\mu \equiv  {1\over 2}\left(-H_\mu-g^{\alpha\beta} g_{\mu\alpha,\beta}
+{1\over 2}g^{\alpha\beta}g_{\alpha\beta,\mu}
\right). 
\end{equation}
where $H_\mu$ are given functions of the coordinates (``gauge source
functions''), we obtain the Einstein equations in the generalised
harmonic gauge. In harmonic gauge all ten $g_{\mu\nu}$, or
equivalently $\g_{ij}$, $\alpha$ and $\beta^i$ obey quasilinear
second-order evolution equations whose principal part is just the wave
operator on the spacetime with metric $g_{\mu\nu}$.

The constraints $Z_\mu=0$ must still be obeyed to obtain a solution of
the Einstein equations, but they are now the harmonic gauge constraints
\begin{equation}
\label{harmonicconstraints}
\Box x^\mu={g^{\mu\nu}}_{,\nu}+{1\over 2}g^{\mu\nu}
g^{\alpha\beta}g_{\alpha\beta,\nu}=H^\mu,
\end{equation}
where $\Box$ is the scalar wave operator \cite{Wald}. In a 3+1 split,
these constraints can be solved for $\dot\alpha$ and $\dot\beta^i$,
and thus constrain the free data for the wave equations for $\alpha$
and $\beta^i$.

The substitution (\ref{H4}) does not change the constraint system
itself at all, which is still the wave equation for $Z_\mu$. The only
difference is now that the $Z_\mu$ are no longer dynamical variables
but are shorthands for the harmonic gauge constraints. We immediately
obtain a prescription for damping all constraints (gauge and Einstein)
in numerical free harmonic evolutions: we modify the Einstein
equations in harmonic gauge to 
\begin{equation}
\label{dampedZ4coords}
R_{\mu\nu}+\nabla_\mu Z_\nu +\nabla_\nu Z_\mu
-\kappa(t_\mu Z_\nu+t_\nu Z_\mu-g_{\mu\nu}t^\lambda Z_\lambda)=0, 
\end{equation}
where $Z_\mu$ is the shorthand (\ref{H4}). (We have used
Greek indices to indicate that these are not tensor
equations but hold only in harmonic coordinates.) As the constraint
system is identical to that of Z4, the same mode analysis applies,
showing that all Fourier modes are damped except the constant in space
modes.

%%%%%%%%%%%%%%%%%%%%%%%%%%%%%%%%%%%%%%%%%%%%%%%%%%%%%%%%%%%%%%%%%%%%%%%%

\section{Z4 and NOR/BSSN}

We obtain the NOR system \cite{NOR} by the
substitutions
\begin{equation}
\label{NOR}
\theta\to 0 , \quad Z_i\to {1\over 2}\left(f_i - \g^{jk}\g_{ij,k} +
\frac{\rho}{2} \g^{jk}\g_{jk,i}
\right)
\end{equation}
in terms of a new variable $f_i$. ($\rho$ is the constant parameter of
the same name in \cite{NOR}, but is different from the $\rho$
introduced in (\ref{Ricci}), which is now set to zero.) The variable
$\theta$ disappears, and $Z_i$ is now the definition constraint of the
new auxiliary variable $f_i$. (Note that $Z_i=-G_i$ in the notation of
\cite{NOR} and \cite{bssn2}.)  Essentially this change of variables
was used in \cite{Z4II} to obtain the BSSN system from the Z4 system,
where constraint damping was also pointed out. The constraints are
$H=0$, $M_i=0$, and $Z_i=0$. The constraint evolution system is
obtained from that of the Z4 system by setting $\theta$ to zero. (The
Z4 system with $\theta=0$ but treating $Z_i$ as dynamical variables is
called the Z3 system, and was proposed in \cite{Z3}. Its constraint
evolution system is identical to that of the NOR system.)

If we repeat the mode analysis for the Z3 system, we find that the
dispersion relation relation for the group of vectors transverse to
$\omega_i$, $(M_A,Z_A)$ is given by
\begin{equation}
s={1\over 2}\left(-\kappa\pm\sqrt{\kappa^2-4\omega^2}\right),
\end{equation}
but for the group of scalars $(H,M_n,Z_n)$ the
modes are given by 
\begin{equation}
s=-\kappa,\pm i\omega,
\end{equation}
so there are two undamped modes. This means we can damp 5 of the 7
constraints of the NOR system (or the BSSN system) by a suitable
modification given essentially by (\ref{dampedZ4coords}) with
(\ref{NOR}).

%%%%%%%%%%%%%%%%%%%%%%%%%%%%%%%%%%%%%%%%%%%%%%%%%%%%%%%%%%%%%%%%%%%%%%

\section{Conclusions}

We have given a $\lambda$ system for general relativity that is
optimal in the sense that the only variables in addition to the usual
ADM variables $\gamma_{ij}$ and $K_{ij}$ are the four $\lambda$
variables $\theta$ and $Z_i$ that are associated with the four
Einstein constraints $H$ and $M_i$. All constraints except the
constant in space modes are damped through lower order terms. This
system is already symmetric hyperbolic (for harmonic slicing and fixed
shift) without the need for other auxiliary variables. It is therefore
a good testbed for investigating the usefulness of $\lambda$ systems
in general relativity in general.

Within the framework of any $\lambda$ system, it is impossible to
obtain damping of homogeneous constraint modes because the damping is
through a friction term. However, it is likely that in the near future
well-posed initial-boundary value problems will be constructed in which the
constraint system implicitly obeys maximally dissipative boundary
conditions. When these are of Dirichlet type, the
homogeneous constraint modes should also be eliminated. 

We have pointed out that the Z4 system is closely related to harmonic
gauge, and that the two share the same constraint evolution
system. Therefore constraint damping works in the same way in harmonic
gauge evolutions, and all 8 constraints can be damped. Similarly, in the
Z3 system and its counterpart the NOR/BSSN system, 5 of 7 constraints
can be damped. 

%%%%%%%%%%%%%%%%%%%%%%%%%%%%%%%%%%%%%%%%%%%%%%%%%%%%%%%%%%%%%%%%%%%%%%%

\acknowledgments

JMM was supported by the Spanish MEC under the research
projects BFM2002-04031-C02-02 and FIS2004-01912. GC was supported by a
Marie Curie Intra-European Fellowship within the 6th Framework. 

%%%%%%%%%%%%%%%%%%%%%%%%%%%%%%%%%%%%%%%%%%%%%%%%%%%%%%%%%%%%%%%%%%%%%%%%

%%%%%%%%%%%%%%%%%%%%%%%%%%%%%%%%%%%%%%%%%%%%%%%%%%%%%%%%%%%%%


\begin{thebibliography}{}

\bibitem{Tiglio} M. Tiglio, Dynamical control of the constraints growth in free
    evolutions of Einstein's equations, e-print gr-qc/0304062, unpublished. 

\bibitem{Calabrese} G. Calabrese, Class. Quant. Grav. {\bf 21}, 4025
  (2004).

\bibitem{Brodbecketal} O. Brodbeck, S. Frittelli, P. Huebner, and
  O. A. Reula, J. Math. Phys. {\bf 40}, 909 (1999).

\bibitem{SiebelHuebner} F. Siebel and P. Huebner, Phys. Rev. D {\bf
  64}, 024021 (2001).
 
\bibitem{Z4III} C. Bona and C. Palenzuela, Phys. Rev. D {\bf 69}, 104003 (2004).
\bibitem{Z4II} C. Bona, T. Ledvinka, C. Palenzuela, and M. Zacek,
  Phys. Rev. D {\bf 69}, 064036 (2004).

\bibitem{Z4I} C. Bona, T. Ledvinka, C. Palenzuela, and M. Zacek,
  Phys. Rev. D {\bf 67}, 104005 (2003).

\bibitem{Z3} C. Bona, T. Ledvinka, C. Palenzuela, and M. Zacek,
  Phys. Rev. D {\bf 66}, 084013 (2002).

\bibitem{Friedrich} H. Friedrich, On the non-linearity of the
  subsidiary systems, preprint gr-qc/0504129. 

\bibitem{Wald} R M Wald, {\it General Relativity}, University of
  Chicago Press, 1980. 

\bibitem{NOR} G Nagy, O E Ortiz and O A Reula, Phys. Rev. D {\bf 70},
044012 (2004)

\bibitem{bssn2} C Gundlach and J M Mart\'\i n-Garc\'\i a, Phys. Rev. D
  {\bf 70}, 044032 (2004).

\end{thebibliography}
\end{document}